\documentclass[showpacs,preprintnumbers,twocolumn,prb]{revtex4}
\usepackage{graphicx}

\usepackage{amssymb}

\begin{document}

\title{Magnetic-dipole induced appearance of vortices in a bilayered superconductor/soft-magnet heterostructure}

\author{S.~V.~Yampolskii}
\altaffiliation[On leave from ]{Donetsk Institute for Physics and Technology,
National Academy of Sciences of Ukraine, 83114 Donetsk, Ukraine. Electronic address:
yampolsk@tgm.tu-darmstadt.de}
\author{G.~I.~Yampolskaya}
\altaffiliation[On leave from ]{Donetsk Institute for Physics and Technology,
National Academy of Sciences of Ukraine, 83114 Donetsk, Ukraine.}
\author{H.~Rauh}
\affiliation{Institut f\"{u}r Materialwissenschaft, Technische
Universit\"{a}t Darmstadt, 64287 Darmstadt, Germany}
\date{3 July, 2006}

\begin{abstract}
The penetration of the magnetic field of an infinitesimal magnetic dipole 
into a bilayered type-II superconductor/soft-magnet heterostructure is studied 
on the basis of the classical London approach. The critical values of the 
dipole moment for the first appearance of a single magnetic vortex and, respectively, 
a magnetic vortex-antivortex pair in the superconductor constituent are obtained, 
when the magnetic dipole faces the superconductor or the soft-magnet constituent. 
This reveals that the soft-magnet constituent inhibits penetration of vortices into the superconductor 
constituent, when the dipole faces the soft-magnet constituent.

\end{abstract}

\maketitle

Heterostructures built from superconducting and
magnetic materials have attracted much attention during the past few years.
This is especially true of heterostructures involving planar superconductors (SCs)
and ferromagnetic dots (FDs) which bring about an enhancement of vortex pinning 
and, concomitantly, an increase of the SC critical current and critical 
field~\cite{rev1,rev2,rev3}. Soft-magnet (SM) constituents offer the possibility 
to greatly improve the performance of SCs by shielding the transport current 
self-induced magnetic field as well as an externally imposed magnetic 
field~\cite{strip1,Campbell,wire1,wire1a,wire2,wire2a}. Here, we study the penetration 
of the magnetic field of a magnetic dipole (MD) of zero extent (deemed to represent 
a small FD) into a bilayered SC/SM heterostructure, when the MD faces the SC or 
the SM constituent. Making recourse to the classical London approach, we compare 
the conditions for the first appearance of a single magnetic vortex and, respectively, 
a magnetic vortex-antivortex pair in the SC constituent.

Let us consider an infinitely extended heterostructure made up of
a type-II SC layer with thickness~$d_S$ and a SM layer with
thickness~$d_M$, and a nearby MD, its moment ${\bf m}$ being directed towards 
the heterostructure.
The relative permeability, $\mu$, is assumed as
the only material characteristic of the SM constituent, apart from the permeability of free space, $\mu_0$.
The condition for the first vortex appearance follows
from the Gibbs free energy of the SC/SM heterostructure due to the presence
of a single vortex, $G_v$, which is the sum of the vortex self-energy and the MD-vortex
interaction energy. Minimization of $G_v$ with respect to the MD-vortex spacing projected onto the SC/SM interface
of the heterostructure and imposition of the requirement $G_{v}=0$ yields the critical
dipole moment $m_{c1}$ provoking the first vortex appearance in the SC constituent.

As has been established before, $m_{c1}$ falls monotonically with increasing
thickness or relative permeability of the SM constituent, when the MD faces the SC constituent~\cite{EPL}. 
This means that the presence of the SM
constituent reduces the critical dipole moment, and hence
facilitates vortex penetration into the SC constituent. By
contrast, when the MD faces the SM constituent, $m_{c1}$ rises monotonically with increasing
thickness or relative permeability of the SM constituent, tending
to a linear increase for large values of the latter quantity; 
behaviour resembling that of a magnetically
shielded cylindrical SC filament~\cite{wire2,wire2a}. Hence, for
this location of the MD, the SM layer can significantly inhibit
vortex penetration into the SC constituent, preserving the latter
in the Meissner state.

It is well known that magnetic vortex-antivortex pairs consisting of a vortex
positioned directly underneath the MD and an antivortex displaced from it, separated by a 
potential barrier due to their interaction with the MD, can also be stabilized 
in the SC constituent~\cite{dip1}.
\begin{figure}[b]
\vspace{-5mm}
\includegraphics[width=8.5cm]{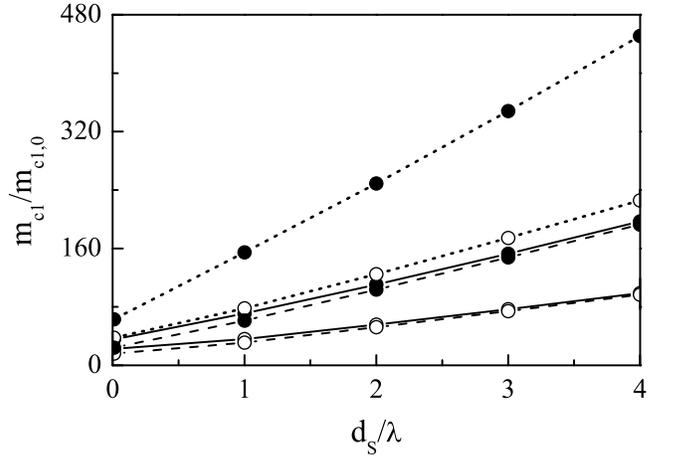}
\caption{Variation of the
normalized critical dipole moment $m_{c1}/m_{c1,0}$ for the first appearance of a single magnetic 
vortex (full circles) and, respectively,  
a magnetic vortex-antivortex pair (open circles) with
the normalized thickness $d_{S}/\lambda$ of the SC constituent of the SC/SM
heterostructure, for the parameter values $d_M/\lambda=0.1$ and $\mu = 50$ of the SM constituent
in conjunction with the prescribed normalized distance 
between the MD and the SC constituent, $l/\lambda=2.5$, 
when the MD faces the SM constituent (dotted lines) or the SC constituent (dashed lines). 
The corrsponding variations in the absence of the SM constituent are shown by solid lines. 
Here, $\lambda$ is the London penetration depth and $m_{c1,0}=\lambda \Phi_0 /\mu_0$ 
with the quantum of magnetic flux, $\Phi_0$, denotes a convenient reference value of the moment of the MD.}
\label{fig1}
\end{figure}
Addressing the Gibbs free energy of 
the SC/SM heterostructure in the presence of such a vortex-antivortex pair, $G_{va}$,
and proceeding in the vein remarked on above, we find that the critical 
dipole moment for the first vortex-antivortex pair
appearance is always larger than that for the first single vortex appearance, 
as demonstrated in Fig.~\ref{fig1}.
Remarkably, this result holds true even in the absence of the SM constituent.
Nevertheless, the SM constituent effects 
the appearance of a vortex-antivortex pair qualitatively in the same manner 
as that of a single vortex: it slightly decreases $m_{c1}$, and hence facilitates 
the appearance of the vortex-antivortex pair, when the MD faces the SC constituent, 
but significantly increaces $m_{c1}$, and hence impedes the creation of a vortex-antivortex pair, 
when the MD faces the SM constituent. 

In conclusion, the presence of the SM constituent can lead to an improvement of 
the superconducting properties of the heterostructure under consideration by 
inhibiting penetration of magnetic vortices into the SC constituent, when 
the MD faces the SM constituent.

\begin{acknowledgments} 
Stimulating discussions with Yu.~A.~Genenko are gratefully acknowledged. 
This study was supported by a research grant from the German Research Foundation (DFG).
\end{acknowledgments}

\end{document}